\documentstyle[12pt]{article}

\begin{document}
\hbox to \hsize{\hfil hep-ph/9610537}
\begin{center}
{\large\bf \vspace{5mm}
 Small $P_{\perp}^2$ Elastic Proton-Proton Polarization near 300 GeV}\\[3ex]
A. D. Krisch$^1$ and \underline{S. M. Troshin$^2$}\\
\vspace{5mm}
{\small \it
(1) Randall Laboratory of Physics, University of Michigan,
Ann Arbor, MI 48108-1120\\
(2) Institute for High Energy Physics,
  Protvino, Moscow Region, 142284 Russia}
\end{center}
\begin{center}
\begin{minipage}{130mm}
\small
Using two hopefully reasonable assumptions plus existing data on the
analyzing power in high energy proton-proton elastic
scattering, we estimate that the hadronic helicity-flip amplitude at
$P^2_{\perp}$ of 0.003 (GeV/c)$^2$ is neglible near 300 GeV.
This would allow
the absolute calibration of a CNI polarimeter.

\end{minipage}
\end{center}
To successfully accelerate a polarized proton beam to a very high energy, one
must somehow measure the beam polarization. Unfortunately, above about
100 GeV, the hadronic spin asymmetries used in most present polarimeters are
small and not well known. Fortunately, in 1948 Schwinger [1]
proposed that the neutrons from neutron-nucleus elastic scattering would
have a large and
energy-independent polarization due to the interference between
the hadronic (Nuclear) and electromagnetic (Coulomb) scattering
amplitudes. This Coulomb-Nuclear Interference (CNI) polarization was
subsequently studied for proton--proton elastic scattering
by  Kopeliovich and Lapidus [2] and later by Buttimore,
Gotsman and Leader [3].
The simplest version of the CNI formula has zero parameters.

However, this simplest version does assume
that the hadronic single-helicity-flip amplitude is
 zero. In using the CNI mechanism for a precise high energy
 polarimeter, one must question the validity of this assumption.
This paper attempts to make a reliable estimate of the CNI
 analysing power $A_n$ near 300 GeV, by trying to set a limit
on the hadronic single-flip amplitude.  There have been previous attempts
to estimate this single-flip amplitude using other techniques [4, 5] and
one attempt to measure the CNI polarization at 200 GeV [6].

The total proton--proton elastic
scattering amplitude in the helicity state $i$ is
\begin{equation}
f_i=f_i^{C}+f_i^{N};
\end{equation}
the electromagnetic amplitudes $f_i^C$ can be easily calculated using
 QED. At small $P^2_{\perp}$ or $-t$, the dominant hadronic amplitudes
$f^N_i$ seem to fit an exponential; its parameters are well determined
by measurements of the total cross--section,
the slope of the diffraction peak, and
the ratio of the real to imaginary amplitudes.

The hadronic part of the proton-proton interaction seems to be
helicity conserving;
the CNI analysing power would be then due to the
interference between a real electromagnetic helicity-flip
amplitude and an imaginary hadronic helicity-nonflip
amplitude. However, the hadronic interaction may not conserve helicity
in small angle
scattering. Helicity conservation
does not follow from QCD in a region where chiral symmetry is spontaneously
broken. In Regge
theory the Pomeron is usually assumed to be helicity conserving;
however this assumption has not been tested.  Moreover
it was shown that unitarity, which is the absorptive correction, generates
different phases in the helicity-flip and nonflip Pomeron
contributions [7].
The CNI analysing power would certainly change if there were any significant
hadronic single-helicity-flip amplitude.  Thus,
we will try to use the existing experimental data on the analysing power
in proton-proton elastic scattering to estimate the hadronic
single-helicity-flip
amplitude near \hbox{$E_{lab}$ = 300 GeV,}
where one might use a CNI polarimeter with a
polarized proton beam [8, 9].

In standard calculations, the CNI analyzing power arises
through the interference between the Coulomb helicity-single-flip
amplitude $f^C_5$
and the hadronic (Nuclear) helicity-nonflip
amplitudes $f_1^N$ and $f_3^N$. However, there could also be a non-zero
hadronic helicity-single-flip
amplitude $f^N_5$, which can be parameterized by
\begin{equation}
\tau (s,t)\equiv\frac{m}{\sqrt{-t}}\frac{|f^N_5|}{|f^N_1|}.
\label{td}
\end{equation}

\noindent We assume that, at small $-t$, the ratio
$\tau$ does not depend strongly on $-t$.  One can then
estimate $\tau$ from the high energy A$_n$
data near $-t$ = 0.2 (GeV/c)$^2$.
\noindent At small $t$, one can  neglect  the three hadronic
helicity-flip amplitudes $f^N_2$, $f^N_4$ and $f^N_5$ in comparison
to the helicity-nonflip amplitudes
$f^N_1$ and $f^N_3$.
Indeed, at small $t$, due to conservation of
angular momentum, $f^N_4$ is proportional
to $-t$ while $f^N_5$ is proportional to $\sqrt{-t}$.
Moreover, at small $t$, the
imaginary part of $f^N_2$ is proportional
 to $\Delta\sigma_t$, which is the difference between the
transverse-spin total cross-sections, \mbox{$\sigma^{tot}$
\footnotesize{$(\uparrow\uparrow$)}
\normalsize{$ - \sigma^{tot}$}
\footnotesize{$(\uparrow\downarrow$)};}
\normalsize
this difference is probably small at high energies
[10].

We shall also assume  that $f^N_1=f^N_3$ at small $t$, since
the existing 12 GeV data on $A_{ll}$ is small,
and the existing $\Delta\sigma_l$ data is small and decreases with energy.
Thus, for purely hadronic processes at small $t$,
\begin{equation}
A_n\frac{d\sigma}{dt}  \simeq
\frac{-8\pi}{s(s-4m^2)}
\mbox{Im} [f^N_1f^{N*}_5],
\label{at}
\end{equation}
\noindent where
\begin{equation}
\quad\mbox\quad
\frac{d\sigma}{dt}  \simeq
\frac{4\pi}{s(s-4m^2)}
|f^N_1|^2.
\label{att}
\end{equation}

Note that, at high energy and small $-t$,
the spin-nonflip amplitude $f^N_1$ is primarily imaginary
because the elastic hadronic nonflip scattering
is primarily the diffractive shadow scattering due to the dominant inelastic
processes; thus, \hbox{$f^N_1 \simeq i|f^N_1|$.}
However, the spin-flip amplitude $f^N_5$ can have both a
real and an imaginary part; writing it as
$f^N_5=|f^N_5|e^{i\phi_5}$,
we have
\begin{equation}
A_n\frac{d\sigma}{dt}\simeq
\frac{-8\pi}{s(s-4m^2)}
|f^N_1||f^N_5|\cos\phi_5,
\label{x}
\end{equation}
 where $\phi_{5}$ is the phase.  Note that $\phi_{5} = \pi/2$
corresponds to a pure imaginary
helicity-single-flip
amplitude while the phases $0$ and $\pi$ correspond to a pure real
$f^N_5$.

Now by combining Eqs. (4) and (5), we obtain
\begin{equation}
A_n=-2 \frac{|f^N_5|}{|f^N_1|}\cos\phi_5
\label{xx}
\end{equation}
 Combining Eqs. (2) and (6), we then
obtain a simple approximation for $\tau$,
which is valid at small $-t$.
\begin{equation}
\tau\cos\phi_5 \simeq -\frac{A_nm}{2\sqrt{-t}}.
\end{equation}

For any hadronic scattering process,
angular momentum conservation implies that  $A_n$
must go  to zero at $t=0$; for the
scattering of two spin--$1/2$ protons,  $A_n$ should be proportional
to $\sqrt{-t}$ near $t = 0$.
Moreover, the experimental data near $-t$ = 0.2 (GeV/c)$^2$
 imply an $s^{-1}$ energy dependence
for $A_n$; therefore, we have
\begin{equation}
A_n\simeq \frac{A\sqrt{-t}}{s}.
\end{equation}
 The constant $A\simeq 3.6$ GeV can be obtained from the $A_n=1.6/s$ fit
to the experimental data.
Then the estimate for
the parameter $\tau$ is
\begin{equation}
\tau\cos\phi_5 \simeq -\frac{Am}{2s} \simeq -\frac{1.7}{s}.\label{tau}
\end{equation}

\noindent At 300 GeV
we then obtain
$\tau\cos\phi_5\simeq 0.003$, which we assumed can be extrapolated from
$-t$ = 0.2 to 0.003 (GeV/c)$^2$.

Now we can estimate the
hadronic contribution to the full single helicity-flip amplitude
$f_5$~=~$f^C_5$+$f^N_5$, where

\begin{equation}
f^C_5\simeq -\frac{\alpha s g F_1(t) F_2(t)}{2m\sqrt{-t}}.
\label{can}
\end{equation}

\noindent where $\alpha = 1/137$ and $g = 1.79$ is the proton's
anomalous magnetic moment, while $F_1(t)$ and $F_2(t)$ are the
proton's form factors.
We use the standard exponential parameterization for
the hadronic amplitude at high energies and small~$t$~[2, 3, 5]

\begin{equation}
f^N_5~=\frac{\tau(\cos\phi_5+i\sin\phi_5)~s \sqrt{-t}~(\rho^2+1)^{1/2}
\sigma_{tot}~e^{bt/2}}{8\pi~m}
\label{sum}
\end{equation}
\noindent At 300 GeV, the ratio of the real to the imaginary forward
scattering amplitude squared $\rho^2$ is experimentally
well below 0.003 [11],
while $\sigma_{tot}$ is about 40 mb.
Moreover, the $e^{bt/2}$ term and both form factors are all
very close to 1 at \hbox{$-t$ = 0.003 (GeV/c)$^2$.}
Thus, at 300 GeV and \hbox{$-t$ = 0.003 (GeV/c)$^2$} we obtain

\begin{equation}
\frac{f^N_5}{f^C_5}
\simeq 1.9\tau(\cos\phi_5 + i\sin\phi_5)
\label{rtt}
\end{equation}

Now we assume that $\tau\sin\phi_5$ is not much larger than $\tau\cos\phi_5$.
This is because the helicity-flip amplitudes certainly can not be diffractive
shadow scattering, since the final and initial spin states are different.
Then we estimate that

\begin{equation}
\tau(\cos\phi_5 + i\sin\phi_5) = 0.003(1+i),
\label{rt}
\end{equation}
thus we have

\begin{equation}
\frac{|f^N_5|}{|f^C_5|}\simeq 5.7 \cdot 10^{-3} |1 + i| < 10^{-2}
\label{rtrt}
\end{equation}

\noindent Thus, at \hbox {$-t=0.003~(\rm GeV/c)^2$} and 300 GeV,
the hadronic amplitude
 is at least 100 times smaller than the Coulomb amplitude.
Therefore, the two assumptions: \\
\vbox{\hspace{0.3in} 1.  $\tau$ depends weakly on $-t$} \\
\vbox{\hspace{0.3in} 2.  $\phi^N_5$ is not pure imaginary} \\
allow one to neglect $f^N_5$
and use the standard calculation of the CNI analysing power [2, 3].
 This result does not
contradict the earlier estimate [5], but it does set a
more stringent limit on the hadronic helicity-flip amplitude.

This work was supported by research grants from the U.S.
 Department of Energy
and the Russian Ministry of Science. One of us (S.T.)
is grateful to the Organizing Committee of SPIN96 Simposium
for financial support.  We would like to thank
N.~E.~Tyurin, \hbox{B.~B.~Blinov,} and the other
members of the SPIN Collaboration
for their help and advice. We are especially
grateful to N. H. Buttimore and T.~L.~Trueman for their comments.

\vspace{0.2cm}
\vfill
{\small\begin{description}
\item{[1]}
J. Schwinger, Phys. Rev. \bf 73\rm , 407 (1948).
\item{[2]}
B. Z. Kopeliovich and I. I. Lapidus, Sov. J. Nucl. Phys. \bf 19\rm ,
114 (1974).
\item{[3]}
N. H. Buttimore, E. Gotsman and E. Leader, Phys. Rev. \bf D18\rm ,
694 (1978).
\item{[4]}
K. Hinotani {\it et al.}, Nuov. Cim. \bf A52\rm , 363 (1979).
\item{[5]}
N. Akchurin, N. H. Buttimore and A. Penzo, Phys. Rev. \bf D51\rm ,
3944 (1995).
\item{[6]}
N. Akchurin {\it et al.}, Phys. Rev. \bf D48\rm , 3026 (1993).
\item{[7]}
S. M. Troshin and N. E. Tyurin, Spin Phenomena in Particle Interactions,
World Sci. Press (Singapore), 1994.
\item{[8]}
SPIN Collaboration A. D. Krisch {\it et al., Acceleration of Polarized
Protons to
120 GeV and 1 TeV}, unpub. Univ. of Michigan Report UM HE 95-09
(July 24, 1995).
\item{[9]}
M. Beddo {\it et al.,} RHIC Spin Proposal, Brookhaven National Laboratory
(1992), (unpublished).
\item{[10]}
W. de Boer {\it et al.}, Phys. Rev. Letts. \bf 34B\rm, 558 (1975).
\item{[11]}
J. P. Burq {\it et al.}, Phys. Lett. \bf 109B\rm , 124 (1982).
\end{description}}
\end{document}